# Bayesian approach to electron correlation in density functional theory

Roumen Tsekov and Georgi N. Vayssilov
Faculty of Chemistry, University of Sofia, 1164 Sofia, Bulgaria

In the present communication we applied the Bayesian conditional probability approach to the wave function of a many-electron system that resulted in appearance of a quantum vector potential in the DFT Schrödinger equation due to electron correlation, apart from the correlation energy term. Mathematically, the effect of this vector potential is equivalent to a magnetic field that corresponds in particular to a conservative irrotational one if it is considered in connection with the correlation potential. An analysis of the effect of the correlation momentum on the electronic transitions suggested that the electron correlation increases the transition probability.

One of the critical points in the quantum chemical methods based either on post Hartree-Fock (HF) approaches or on density functional theory (DFT) is connected with proper accounting for the electron correlation[1,2,3]. In many-electron systems this correlation has crucial short- and long-range effects; it influences the electronic state of the system[4] and contributes to the intra- and inter-molecular van der Waals interactions.[5,6,7] Traditionally, the electron correlation is considered to affect the total electronic energy of the system and is measured by the correlation energy[8] defined as the decrease of the total energy with respect to the energy obtained with the HF approximation. The correlation energy is estimated by different types of multi-determinant post HF methods,[9,10] e.g. configuration interaction or coupled cluster methods, by the Hartree-Fock-Bogolyubov approach or by correlation component of the exchange correlation functional within the frame of DFT.[11,12,13] In the present communication using the Bayesian probability approach we have shown that the electron correlation modifies not only the potential energy terms in the Schrödinger equation but is also responsible for a vector potential in this equation which originates from the electron kinetic energy operator. The effects of this vector potential in some simplified cases are discussed. In particular, the vector potential is equivalent to a collective magnetic-like field. Hence, the latter possesses no monopole carriers, similar to magnetic interactions.

Let us consider now an atom, molecule or body possessing N electrons. The stationary Schrödinger equation for the N-electron wave function $\Psi$ reads

$$\sum_{j=1}^{N}(\hat{p}_j^2/2m + U_j + \sum_{k=j+1}^{N} V_{jk})\Psi = E\Psi \tag{1}$$

where $\{\hat{p}_k \equiv -i\hbar \nabla_k\}$ are the momentum operators of the electrons with mass $m$, $\{U_k\}$ are the potentials of the electron-nuclei interactions, $\{V_{jk}\}$ are the potentials of the electron-electron interactions, and $E$ is the total electronic energy. The wave function can be presented generally in a Bayesian form[14,15,16,17,18]

$$\Psi(r_1,\cdots,r_N) = \phi(r_2,\cdots,r_N | r_1)\psi(r_1) \tag{2}$$

where $\psi$ is a single electron wave function and $\phi$ is the conditional wave function of the remaining N-1 electrons. The square of the latter represents the conditional probability to find a configuration of the remaining electrons if the first electron is fixed at the point $r_1$. The many-particle function $\phi$ bears all the necessary symmetric properties of the wave function $\Psi$. In the present paper spin variables are not considered; they are solely applied via the Pauli Exclusion Principle to orbital occupation. The presentation (2) can be straightforward transformed to a density type representation since according to the Bogolyubov hierarchy[19] the dependence of the conditional probability density on $r_1$ goes principally via the single electron density $\rho = |\psi|^2$, i.e. the conditional wave function $\phi(r_2,\cdots,r_N | r_1) = \phi(r_2,\cdots,r_N | \psi(r_1))$ is a functional of $\psi$. This presentation, which reflects in a density functional theory, is our basic innovation to Hunter's seminal papers from the previous century.

Introducing Eq. (2) in Eq. (1), multiplying the resulting equation by the complex-conjugated conditional wave function and integrating along the coordinates of the remaining N-1 electrons lead to

$$\hat{p}_1^2\psi/2m + \langle\phi|\hat{p}_1|\phi\rangle\hat{p}_1\psi/m + \sum_{j=1}^{N}\langle\phi|\hat{p}_j^2/2m + U_j + \sum_{k=j+1}^{N}V_{jk}|\phi\rangle\psi = E\psi \tag{3}$$

From the obvious consequence $\hat{p}_1\langle\phi|\phi\rangle = 0$ of the conditional wave function normalization $\langle\phi|\phi\rangle = 1$ it follows that the term $p_{corr} \equiv -\langle\phi|\hat{p}_1|\phi\rangle$ is a real function of $r_1$. It possesses a pure correlation origin since in the case of independent electrons the conditional wave function $\phi$ does not depend on $r_1$ and, hence, $p_{corr} = 0$. Introducing this notation, Eq. (3) can be rewritten in the form

$$(\hat{p}_1 - p_{corr})^2\psi/2m + (U_1 + V_H + W_{corr})\psi = \varepsilon_1\psi \tag{4}$$

where $U_1$ is the Coulomb electrostatic energy of interaction of the target electron by the nuclei, $V_H$ is the Hartree potential describing uncorrelated electron-electron interactions, $\{\varepsilon_j\}$ are the single electron energies ($E = \sum \varepsilon_j$), and the correlation potential

$$W_{corr} \equiv \langle \phi | (\hat{p}_1 + p_{corr})^2 | \phi \rangle / 2m + \langle \phi | \sum_{k=2}^{N} V_{1k} - V_H | \phi \rangle + \sum_{j=2}^{N} \langle \phi | \hat{p}_j^2 / 2m + U_j + \sum_{k=j+1}^{N} V_{jk} - \varepsilon_j | \phi \rangle \quad (5)$$

is a real function of $r_1$. As is seen, all non-classical interactions are collected fully in $W_{corr}(r_1)$, which consists of three distinguished terms: kinetic, electrostatic and entropic ones. The latter represents the energy changes of the separate N-1 electron system, influenced by the target first electron merely via the conditional distribution density.[18] The correlation term $W_{corr}(\rho)$ plays the role of an effective DFT exchange scalar potential, while the correlation momentum $p_{corr}(\rho)$ is an effective DFT vector potential. The key feature of Eq. (4) is the appearance of correlation effects not only in as component of the potential energy but also in the vector potential. Mathematically, the effect of the vector potential $p_{corr}$ is equivalent to a magnetic-like field. This analogy suggests that the correlation vector potential induces circular motion of the electrons and slows them down. Note that the gradient-corrected exchange-correlation functional in DFT[11,12,13] already includes gradients of the electron density but due to different reasons.[20]

Following the causal Onsager flow-force relationships from the non-equilibrium thermodynamics, it seems reasonable to accept that the correlation momentum $p_{corr}$ is due to the correlation force $-\nabla_1 W_{corr}$, since they vanish simultaneously at large distance between the electrons. The correlation momentum describes an effective friction force $\gamma p_{corr}$, where $\gamma$ is a specific friction coefficient among electrons.[21] In order the whole system to be stationary it is required that this force is compensated by the correlation one, which provides the linear relationship $\gamma p_{corr} = -\nabla_1 W_{corr}$. Thus, in this case $p_{corr}$ corresponds to an irrotational vector field, i.e. $\nabla_1 \times p_{corr} = 0$. Introducing in Eq. (4) of the expression $p_{corr} = -\nabla_1 W_{corr} / \gamma$ one can rewrite it in the form

$$\hat{p}_1^2 \varphi / 2m + (U_1 + V_H + W_{corr}) \varphi = \varepsilon_1 \varphi \quad (6)$$

where the new wave function is defined by $\varphi \equiv \psi \exp(iW_{corr} / \hbar \gamma)$. Equation (6) represents an ordinary Kohn-Sham-like equation and providing a model for the correlation potential $W_{corr}$ one can solve it via traditional ways. What is important is that the complete wave function should be reconstructed from this solution via

$$\psi = \varphi \exp(-iW_{corr}/\hbar\gamma) \tag{7}$$

by the use of the employed model for $W_{corr}$ and the 'empirical' coefficient $\gamma$. It is significant to note that the transformation (7) affects neither the electron density $\rho = |\psi|^2 = |\varphi|^2$ nor the energy spectrum $\{\varepsilon_{1,n}\}$ of the system, which can be calculated from Eq. (6). Moreover, the following laws of conservation of momentum and energy hold as well

$$\langle \psi | \hat{p}_1 | \psi \rangle = \langle \varphi | (\hat{p}_1 + p_{corr}) | \varphi \rangle \qquad \langle \psi | (\hat{p}_1 - p_{corr}) | \psi \rangle = \langle \varphi | \hat{p}_1 | \varphi \rangle$$

$$\langle \psi | \hat{p}_1^2 | \psi \rangle = \langle \varphi | (\hat{p}_1 + p_{corr})^2 | \varphi \rangle \qquad \langle \psi | (\hat{p}_1 - p_{corr})^2 | \psi \rangle = \langle \varphi | \hat{p}_1^2 | \varphi \rangle$$

The effect of the exponential term in Eq. (7) is simply a shift of the wave function phase, which could cause an Aharonov-Bohm effect,[22] being related in general to superconductive currents. The quantum phase plays an important role in quantum systems and for this reason the classical gauge transformation is not applicable. The correlation momentum has no impact on classical systems according to the Bohr–van Leeuwen theorem. Hence, the effect of $p_{corr}$ is expected to affect the electronic transitions via off-diagonal terms in the Fermi golden rule, Weisskopf-Wigner approximation, Landau-Zener formula, etc. For instance, using Eq. (7) the standard oscillator strength can be written in the form

$$\begin{aligned} f_{nm} &\equiv (2m/3\hbar^2)(\varepsilon_{1,m} - \varepsilon_{1,n}) \langle \psi_m | r_1 | \psi_n \rangle^2 \\ &= (2m/3\hbar^2)(\varepsilon_{1,m} - \varepsilon_{1,n}) \langle \varphi_m | r_1 \exp[i(W_{corr,m} - W_{corr,n})/\hbar\gamma] | \varphi_n \rangle^2 \end{aligned} \tag{8}$$

If the phase shift is small the exponential term can be expanded in series to obtain

$$\exp[i(W_{corr,m} - W_{corr,n})/\hbar\gamma] \approx 1 + i(W_{corr,m} - W_{corr,n})/\hbar\gamma \tag{9}$$

which introduced in Eq. (8) results in the approximation

$$f_{nm} = (2m/3\hbar^2)(\varepsilon_{1,m} - \varepsilon_{1,n})[\langle \varphi_n | r_1 | \varphi_m \rangle^2 + \langle \varphi_n | r_1 (W_{corr,m} - W_{corr,n})/\hbar\gamma | \varphi_m \rangle^2] \tag{10}$$

Hence, the correlation effect increases the transition probability, as expected since the correlations increase the wave functions overlap. For instance, the simple Ohmic expression[18] for the correlation momentum $p_{corr} = m\gamma r_1$ corresponds to effective harmonic correlation repulsion

$W_{corr} = -m\gamma^2 r_1^2 / 2$ between electrons, trying to expand the electronic shell. Considering an ionization process with $\varepsilon_{1,\infty} = 0$ and $W_{corr,\infty} = 0$ Eq. (10) acquires the form

$$f_{n\infty} = -(2m/3\hbar^2)\varepsilon_{1,n}[\langle \varphi_n | r_1 | \varphi_\infty \rangle^2 + (m\gamma/2\hbar)^2 \langle \varphi_n | r_1^3 | \varphi_\infty \rangle^2] = f_{n\infty}(\gamma = 0)[1 + (m\gamma r_a^2 / 2\hbar)^2] \quad (11)$$

The characteristic radius $r_a \equiv (\langle \varphi_n | r_1^3 | \varphi_\infty \rangle / \langle \varphi_n | r_1 | \varphi_\infty \rangle)^{1/2}$ is of the order of the atomic radius. The term $2m\gamma r_a^2 / \hbar$ can be interpreted as the ratio between the frequencies of the electron-electron 'collisions' $\gamma$ and of the electron orbital rotations $\hbar / 2mr_a^2$ in the atom or, alternatively, as the ratio between the atomic radius and the electron mean-free-path $\lambda = \hbar / 2mr_a\gamma$. Using the standard formula $\lambda = r_a^3 / 3r_e^2 N$ where $r_e$ is an effective radius for the electron-electron collisions, Eq. (11) can be rewritten in the form $f_{n\infty}(\gamma) / f_{n\infty}(0) = 1 + (3Nr_e^2 / 4r_a^2)^2$. Thus, the correlations between the electrons assist the ionization process and the oscillator strength increases by square on the number of electrons in the atom. An estimate for the effective specific friction coefficient is $\gamma = \hbar / 2mr_a\lambda = 3N\hbar r_e^2 / 2mr_a^4$.

Alternatively, in accordance to the Bogolyubov ansatz[19] the correlation momentum can be expressed by the electron density gradient via $p_{corr} \equiv \langle \phi | i\hbar \nabla_1 | \phi \rangle = \langle \phi | \partial_\rho | \phi \rangle i\hbar \nabla_1 \rho$, where the new statistical moment can be roughly estimated as $\langle \phi | \partial_\rho | \phi \rangle = i / \rho$. The corresponding vector field $p_{corr} = -\hbar \nabla_1 \ln \rho$ is conservative indeed and, in the case of a Gaussian density $\rho$, it is proportional to $r_1$ as modeled before. The mean value of $p_{corr}$ is zero, while its dispersion equals to the Fisher information. The correlation electric current $j_{corr} \equiv e\rho p_{corr} / m = -(e\hbar / m)\nabla_1 \rho$ is diffusive. The vector potential $p_{corr} = -\hbar \nabla_1 \ln \rho$ corresponds to a logarithmic scalar correlation potential $W_{corr} = \hbar\gamma \ln \rho$,[23] which is also purely entropic, reflecting the Shannon information density. Considering the latter as Boltzmann entropy[18] one can interpret the term $\hbar\gamma / k_B$ as an effective temperature of colliding electrons, e.g. $\gamma = k_B T / \hbar$ according to the Eyring-Polanyi equation. The previously introduced harmonic model $W_{corr} = -m\gamma^2 r_1^2 / 2$ corresponds now to $W_{corr} = \hbar\gamma \ln \rho$ with a normal probability density. It is known that a logarithmic nonlinearity in the Schrödinger equation leads to appearance of Gaussian solitons,[24] which are of interest to the contemporary quantum chemistry, physics and information theory. In the case of $W_{corr} = \hbar\gamma \ln \rho$ the wave function from Eq. (7) reduces simply to a rescaling ratio $\psi = \varphi / |\varphi|^{2i}$ with imaginary fractal dimension.

As the only rigorous example one can mention here the Hooke atom (harmonium), which possesses two electrons interacting with the core via harmonic potentials. In this case the Schrödinger equation is analytically solved and the corresponding ground-state wave function reads[25] (the coordinates are presented in atomic units)

$$\Psi = \pi^{3/4}(1+|r_1 - r_2|/2)\exp(-r_1^2/4 - r_2^2/4)/(8+5\pi^{1/2})^{1/2} \qquad (12)$$

It is easy to calculate from this equation the single electron density $\rho$, which is related to single-particle wave function via $\psi = \rho^{1/2}$. Thus, one can calculate analytically the conditional wave function $\phi(r_2|r_1) = \Psi(r_1,r_2)/\psi(r_1)$ and the corresponding $p_{corr}$ and $W_{corr}$ in this simple case. As was shown,[25] however, the correlation corrections in helium are negligible and the Hartree-Fock method works well. We expect important contributions of electron correlations in many-electron systems, which are, however, difficult to solve and require phenomenological models as shown above.

In conclusion, the paper presents an approach to the many-electron problem based on decomposing the many-electron wave function into a conditional probability and marginal probability; the square of the latter represents the one-body density. We have shown that the accounting for the electron correlation via straight forward application of the Bayesian conditional probability to the wave function of the system resulted in appearance of a vector potential in the Kohn-Sham equation. Hence, the correlation is represented not only by a scalar potential, in usual approaches, but also by a vector potential. The effect of this vector potential is mathematically equivalent to a magnetic-like field that corresponds to a solenoidal vector field if it is considered in relation to the correlation energy. Additional analysis of the effect of the correlation momentum on the electronic transitions suggested that the electron correlation increases the transition probability.

In general, if the full N-electron wave function $\Psi$ is known one can calculate the conditional (N-1)-electron function $\phi$ exactly. However, due to obvious difficulties to determine $\Psi$ this program is working only for simple applications, an example of which is mentioned above. As it was demonstrated in the present paper the *ad hoc* introduction of the conditional wave function $\phi$ results in appearance of new statistical moments, which could be anticipated from general perspective. The situation is the same as in statistical physics, where the Liouville equation contains the full description but it is impossible to solve it. Introducing some statistical hypotheses one can derive the Boltzmann or the Fokker-Planck equations from it, which are simpler mathematical problems. Furthermore, one can derive the whole hydrodynamics. Therefore, our analytical procedure of simplification can lead to practical computational methods for many electron systems. The reduced density matrix functional approach is another possible way to describe many electron systems. The advantage of our method is due to use of the conditional wave function $\phi$. Thus, in the case of weak correlations one can easily recognize the physical meaning of the new terms. After some simple assumptions we arrived to a Kohn-Sham-like equation (6) and hence one can employ the standard mathematical tools developed in DFT. The reduced density matrix method is closer to statistical mechanics and gives more coarse-grained picture of the system. Note that the mathematical apparatus of the wave functional quantum mechanics is much more elaborated than that of quantum statistical mechanics.